\documentclass[aps,prl,showpacs,twocolumn,superscriptaddress]{revtex4}
\usepackage{subfigure}
\usepackage{amsmath}
\usepackage{graphicx}
\usepackage{rotating}
\usepackage{epstopdf}

\usepackage{bm}
\usepackage{graphicx}
\usepackage[german,english]{babel}

\def\bbbc{{\mathchoice {\setbox0=\hbox{$\displaystyle\rm C$}\hbox{\hbox
to0pt{\kern0.4\wd0\vrule height0.9\ht0\hss}\box0}}
{\setbox0=\hbox{$\textstyle\rm C$}\hbox{\hbox
to0pt{\kern0.4\wd0\vrule height0.9\ht0\hss}\box0}}
{\setbox0=\hbox{$\scriptstyle\rm C$}\hbox{\hbox
to0pt{\kern0.4\wd0\vrule height0.9\ht0\hss}\box0}}
{\setbox0=\hbox{$\scriptscriptstyle\rm C$}\hbox{\hbox
to0pt{\kern0.4\wd0\vrule height0.9\ht0\hss}\box0}}}}
\newcommand{\beq}{\begin{eqnarray}}
\newcommand{\eeq}{\end{eqnarray}}

\newcommand{\beqa}{\begin{eqnarray}}
\newcommand{\eeqa}{\end{eqnarray}}

\begin{document}

\title{Superconductivity in Ca-intercalated bilayer graphene }
\author{I.I. Mazin}
\affiliation{Code 6393, Naval Research Laboratory, Washington, D.C. 20375}
\author{A. V. Balatsky}
\email{avb@lanl.gov}
\affiliation{Theoretical Division, Los Alamos National Laboratory, Los Alamos, New Mexico
87545, USA}
\affiliation{Center for Integrated Nanotechnology, Los Alamos National Laboratory, Los
Alamos, New Mexico 87545, USA}
\date{Printed \today }

\begin{abstract}
Recent observation of proximity effect \cite{Morpurgo:2007} has
ignited interest in supercondcutivity in graphene and its
derivatives. We consider Ca-intercalated graphene bilayer and argue
that it is a superconductor, and likely with a sizeable $T_{c}$. We
find substantial and suggestive similarities between Ca-intercalated
bilayer (C$_{6}$CaC$_{6}$), and CaC$_{6} $, an established
superconductor with $T_{c}$ = 11.5 K. In particular, the nearly free
electron band, proven to be instrumental for superconductivity
in intercalated graphites, does  cross the chemical potential in (C$_{6}$CaC$%
_{6}$), despite the twice smaller doping level, satisfying the so-called
\textquotedblleft Cambridge criterion\textquotedblright . Calculated
properties of zone-center phonons are very similar to those of CaC$%
_{6}.$ This suggests that the critical temeperature would probably be on the
same scale as in CaC$_{6}$.
\end{abstract}

\pacs{Pacs Numbers: }
\maketitle

The graphite becomes superconducting after intercalation with alkali
elements, with the transition temperature ranging from below 1K for KC$_{8}$
to 11.5 K for CaC$_{6}$ \cite{Belash:1989,Weller:2005,Emery:2005,Hannay:1965}%
. Mechanism of superconductivity seems to be consistent with phonon mediated
pairing \cite{Mazin:2007}.
These graphite
intercalated compounds (GIC) open a promising route to an alternative class
of superconducting of materials with tunable properties.

The recent discovery of graphene\cite{Novoselov:2005,Zhang:2005}, a
single sheet of carbon atoms, has naturally raised a question of
superconductivity in graphene\cite{Neto}. Proximity induced
superconductivity in pure graphene has been demonstrated \cite{Morpurgo:2007}%
. This work shows the potential for developing new superconducting
devices starting with 2D graphene as a basis material.

The purpose of this Letter is to investigate routes of doping
graphene with Ca to an extent that would induce intrinsic
superconductivity. In doing that, we will heavily rely on insights
from now well understood superconductivity in intercalated
graphites.

Obviously, for a large number of graphene layers we would have to recover
bulk superconductivity of CaC$_{6}$. Hence, if superconducting state is
indeed possible, one can intercalate multilayer
compounds with intermediate numbers of layers
and investigate the crossover between 2D
superconducting state in a few layers and the increasingly 3D character of
multilayer compositions. This bottom up approach leads to a reasonable
question to ask: is an intercalated graphene bilayer a superconductor in
itself? We will argue that the answer to this question is affirmative.

\begin{figure}[htb]
\begin{center}
\includegraphics[width=6cm ] {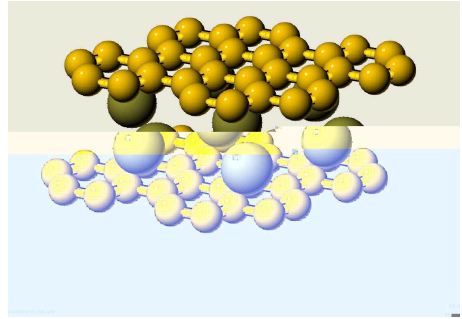}
\end{center}
\caption{ Crystal structure of Ca C6 bilayer is shown. Ca atom sits in the
approximately in the middle of one of hexagons of $C_{6}$. }
\label{CaC6}
\end{figure}

Let us recall the basics of superconductivity in
CaC$_{6}$.\cite{Mazin:2007} It was proposed at a very early stage
that (i) soft Ca modes contribute substantially to electron-phonon
coupling\cite{Mazin:2005} and (ii) a nearly free-electron
three-dimensional electronic band, ``$\zeta$'', an analogue of the
free electron $s-$band in Ca metal, plays an indispensable role in
superconductivity\cite{Csanyi:2005}. An important observation was
made by Littlewood and collaborators \cite{Csanyi:2005} who pointed
out that for all known superconducting GIC a nearly free electron
band (which is well above the Fermi level in the pure graphite)
crosses the Fermi level (the \textquotedblleft Cambridge
criterion\textquotedblright ). Detailed calculations confirmed both
conjectures\cite{Mauri:2005,Sanna} and found
that the Ca phonons provide about half of the coupling strength \cite%
{Hinks:2007}, while electrons experiencing the strongest pairing interaction
are those in the nearly free electron band, although the carbon electrons
also show a sizeable coupling with phonons.

This naturally suggests Ca as a dopant for graphene. We focus on the bilayer
graphene case as a first compound that can be truly intercalated. \cite%
{comment2}. Moreover, intercalation with Ca should provide additional
rigidity, making the new superconductor structurally robust. It is not
obvious, however, that intercalating a bilayer will be as effective as
intercalating the bulk graphite. As a very minimum, intercalating a bilayer
to the same degree as graphite provides twice less carriers, $e.g.,$
C$_{6}$CaC$_{6}$ $vs.$ CaC$_{6}$. \cite{Comment1}. Therefore, the first
question to ask is whether the emprirical \textquotedblleft Cambridge
rule\textquotedblright\ holds, that is, whether the Ca-derived
nearly-free-electron band does cross the Fermi level? Note that as opposed
to the graphites, where this band is three-dimensional and its density of
states depends on filling, in bilayer graphene it is 2D and therefore has
filling-independent density of states (as long as it crosses the Fermi
level). The other point is to investigate how similar are the elastic properties
of Ca intercalated between two graphene sheets to that of the Ca in
graphite. In what follows we will answer both questions affirmatively, based
on first principle calculations, and will therefore suggest that
Ca-intercalated bilayer graphene should be a superconductor with a critical
temperature comparable to that of CaC$_{6}.$

\begin{figure}[htb]
\begin{center}
\includegraphics[width=0.95\linewidth]{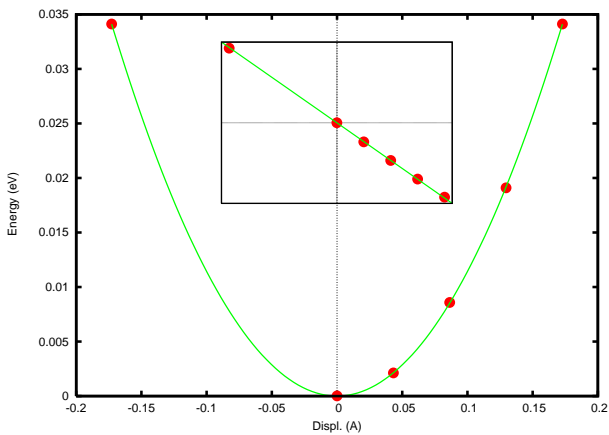}
\end{center}
\caption{ Frozen-phonon energy for a $E_{1u}$ mode calculated by displacing
Ca along the $x$ direction. The inset shows calculated forces acting upon Ca
for the same displacements. The lines are, respectively, the least square
quadratic and linear fits. The coefficients agree within 6\% and yield for
the phonon frequences
of 120 (123) cm$^{-1}$. (Color online)}
\label{phonon}
\end{figure}

To this end, we have used the standard full-potential linear augmented wave
method for band structure calculations\cite{Wien2k} in conjunction with the
density functional theory in a generalized gradient approximation. The
setup and technical details have been described elsewhere\cite{Mazin:2005}.
To imitate an isolated C$_{6}$CaC$_{6}$ trilayer we used an 18.2 \AA\ thick
slab and optimized the distance between the C layers. The latter came out to
be 4.64 \AA , only slightly expanded compared to that in the bulk CaC$_{6},$
namely. This, by itself, indicates that bonding is similar and the phonon
property will probably also be similar. To verify that latter conjecture we
have performed frozen phonon calculations of a Ca $E_{1u}$ phonon mode,
corresponding to the Ca layer sliding with respect to the C layers. The
results are shown in Fig. \ref{phonon}.

Note that the mode is very harmonic. The calculated frequency is 123
cm$^{-1} $ from the total energy fit, and 120 cm$^{-1}$ from the
forces, indicating a good convergence. This frequency is similar to
the corresponding frequency in the bulk CaC$_{6}$\cite{Mauri:2005},
$\approx 115$ cm$^{-1}$, supporting our conjecture that the phonon
properties of a Ca layer sandwiched between two C layers are very
similar to those of Ca in CaC$_{6}$. This result was computed for
purely Ca displacements; allowing for the weak hybridization with C
mode of the same symmetry would bring the frequency slightly down,
making it even closer to that in CaC$_{6}$\cite{Mauri:2005}. We have
also calculated the two $A_{1g}$ modes: the in-plane C mode, that is
expectedly hard, 1370 $cm^{-1}$, and the one that corresponds to
breathing between the C planes. The latter mode is absent in the
bulk compound. It appears to be very soft, 108 $cm^{-1}$. Given that
the position of the free-electron band strongly depends on the
interlayer distance, we assume it will interact strongly with that
band and thus open an additional electron-phonon coupling channel,
absent in CaC$_6$.

\begin{figure}[t]
\begin{center}
\includegraphics[width=0.95\linewidth]{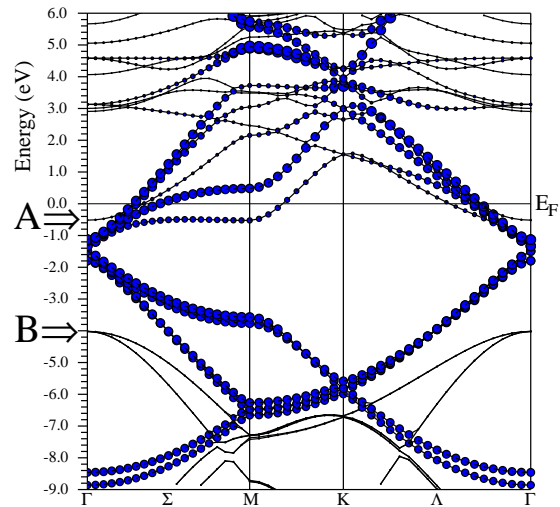}
\end{center}
\caption{ Band structure of the bilayer graphene Ca$_6$CaC$_6$. The size of
the symbols indicates the relative C-$\protect\pi$ character of the
electronic states. The arrow ``A'' points to interlayer nearly-free-electron
$\protect\zeta$ band (see text) and the arrow ``B'' to the bonding $\protect%
\sigma$ band.(Color online). }
\label{spaghetti}
\end{figure}

\begin{figure}[t]
\begin{center}
\includegraphics[width= 0.95\linewidth]{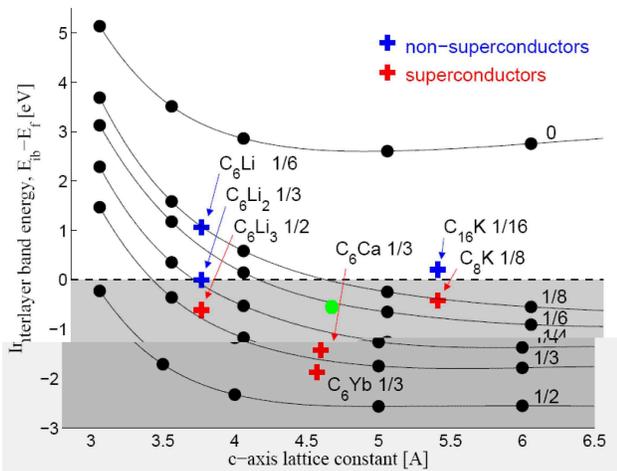}
\end{center}
\caption{ Correlation between the position of nearly free band, brought down
to chemical potential due to Ca and superconducting properties of different
materials, from \protect\cite{Csanyi:2005}. We added to this figure a point
(green circle) representing C$_{6}$CaC$_{6}$ ($cf.$ Fig.
\protect\ref{spaghetti}).(Color Online).}
\label{Littlewood}
\end{figure}

Let us now turn to the electronic properties. In Fig.
\ref{spaghetti} we show the calculated bands with C $p_{z}$
character emphasized (the so-called $\pi $ bands). Obviously, the
Dirac points are now well below the Fermi level. More importantly,
we find the Ca-derived nearly free electron band crossing the Fermi
level with its bottom located at the $\Gamma $ point $0.5$ eV below
the Fermi level. If we place the intercalated bilayer graphene on
the
\textquotedblleft Cambridge plot\textquotedblright\ from Ref. \cite%
{Csanyi:2005} we observe it to be located above CaC$_{6},$ partially due to
the nominally twice smaller doping, 1/6 $vs.$ 1/3, see Fig. \ref{Littlewood}%
.

\begin{figure}[htb]
\begin{center}
\includegraphics[width=0.95\linewidth]{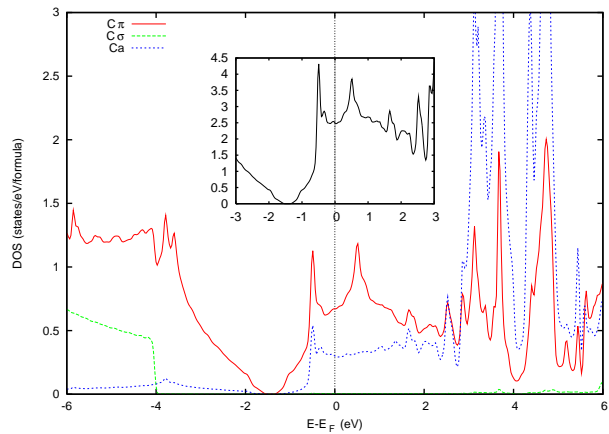}
\end{center}
\caption{ Density of states projected onto individual atomic
functions. Note the sharp onset of the bonding C $\protect\sigma$
bands (long-dash green line) below $-4$ eV and the onset of the $\protect%
\zeta$ band (short-dash blue line) above $-0.5$ eV. The inset shows
the total DOS, which is very close to that of the bulk CaC$_6$ if
compared on the per-C basis. (Color online). } \label{DOS}
\end{figure}

\begin{figure}[htb]
\begin{center}
\includegraphics[width=0.95\linewidth]{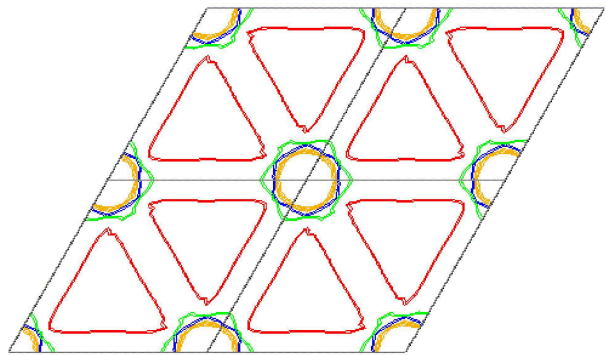}
\end{center}
\caption{ 2D Fermi surface of C$_{6}$CaC$_{6}$. Note a perfect circle
around the $\Gamma$ point (orange online), derived from the $\zeta$
band, and the triangular Fermi contours derived from the Dirac electrons.
(Color online). }
\label{FS}
\end{figure}

Other characteristics of the calculated electronic structure are
shown in Fig. \ref{DOS} (density of states) and \ref{FS} (Fermi
surface). Several observations are in order. First, the density of
states per carbon at the
Fermi level is very similar to that of the bulk CaC$_{6}:$ 2.5 states per C$%
_{12}$ and compared to 1.5 per C$_{6}$ in CaC$_{6}$\cite{Mazin:2005}.
Second, compared to CaC$_{6}$ (see Ref. \cite{Mazin:2005}, Fig. \ref{FS}),
the Fermi level appears in a minimum of the density of states between two
large peaks at $\sim \pm 0.5$ eV, a feature favorable for crystal stability.
Finally, it is worth noting that the Ca-projected DOS is at least half of
the C-projected one. The former comes predominantly from the interlayer
band. Since the LAPW program projects the DOS onto the muffin-tin spheres, and
the interstitial space in this structure is huge, this indicates that the
DOS of the interlayer band is comparable with that of the C-$\pi $ bands,
despite the fact that the corresponding Fermi surface (small orange circle
around the $\Gamma $ point in Fig. \ref{FS}) is so small compared to the
Fermi surfaces of CaC$_{6}$ and YbC$_{6}$ (Refs. \cite{Mazin:2007,Molodtsov}%
). This is a manifestation of the two-dimensional character of the
band structure and independence of the DOS (and the corresponding
electron-phonon coupling\cite{Dolgov}) of the doping level. A
useful, albeit elementary, excersise is to compare DOS of a 3D
parabolic band at a given filling with that of a 2D parabolic band
with the same mass.  It is easy to see that the latter is larger as
long as the Fermi vector in the 3D band $k_{F}<\pi /c$, where $c$ is
the interlayer spacing in the 3D system (geometrical meaning: as
long as the Fermi sphere diameter is smaller than the interplanar
distance in the reciprocal space). Note that in CaC$_{6}$ and
YbC$_{6}$ this condition is $not$ satisfied, and therefore going
from 3D to 2D is beneficial for superconductivity.

As discussed, the coupling of the interlayer band with phonons is
essentially doping-independent (as shown in Ref. \cite{Dolgov}, this holds
even when the FS is so small and the Kohn anomaly so strong that the phonon
self-energy should be calculated self-consistently including the feedback
effects). Based on this, we suggest that pairing coupling constant in Ca$_{6}
$CaC$_{6}$ is similar to the parent 3D compound. However, this does not mean
that additional doping does not help in terms of enhancing the coupling. As
we know from the CaC$_{6}$ calculations, about half of the total coupling
comes from the C $\pi $ bands, and these, being strongly non-parabolic (a
triangular shape of the largest Fermi surface in Fig. \ref{FS} attests to
the fact that the C bands are still fairly close to the Dirac dispersion),
do show energy-dependent DOS, and, by implication, energy-dependent electron
phonon coupling. Adding surface Ca atoms to the intercalated bilayer
graphene has potential to further increase the coupling and enhance the
critical temperature.

Our observation about the nearly free electron band crossing the Fermi level
(Figs. \ref%
{spaghetti},\ref{Littlewood}) and the similarity of the  phonon
spectra in the intercalated bilayer and in the bulk CaC$_6$ are the
main results of this Letter. Based on these similarities,  we
conjecture that Ca intercalated graphene bilayer is a
superconductor.

A few comments are in place regarding the practicality of these
materials. We know now that the single and the bilayer graphenes are
electronically
inhomogeneous, as was seen by scanning probes \cite{Stroscio:2007,Martin:2008}%
. Effects of charge inomogeneity on the superconducting state would
need to be addressed in details if indeed superconductivity is
observed in these materials. Variations of local charge density in
bilayer graphene would play a role of nonmagnetic impurities. As
such, they would be subject to the Anderson theorem and thus would
not be pair-breaking for the isotropic or nearly-isotropic s-wave
superconductivity, as it is believed to be the case in
CaC$_{6}$\cite{Sanna}  and presumably is in C$_{6}$CaC$_{6}$.
Another important effect left outside of the scope of this letter is
the role of substrate. We assumed that substrate effects would be
the strongest for a single layer. One would need to model the
substrate in a manner consistent with the DFT approach used here to
address its role specifically.

In conclusion, we propose to search for superconductivity in a Ca
intercalated graphene bilayer. Our estimates of the phonon frequencies and
possible electron-phonon coupling constant make this material a plausible
candidate for superconductivity with $T_{c}$ in the range of few K, and
possibly above 10 K. We start with an undoped bilayer graphene that is
nonsuperconducting. Ca intercalation renders a band structure that is rather
close to that of the bulk CaC$_{6}$, a known superconductor with $T_{c}=11.5K$.\cite%
{Belash:1989,Weller:2005,Emery:2005,Hannay:1965}. Frequencies of zone center
phonons are very close to those in CaC$_{6}.$ The nearly free electron band
crosses the Fermi level in C$_{6}$CaC$_{6}$, just as it does in all
known superconducting intercalated graphites\cite%
{Csanyi:2005}, thus satisfying the \textquotedblleft Cambridge
conjecture\textquotedblright , Fig(\ref{Littlewood}). Despite the lower
doping level, the 2D character of this band provides an even higher density
of states than in CaC$_{6}.$

The path to design superconducting material proposed here is a
bottom up approach, similar to the multilayer superconducting films.
If the ideas presented here turn out to be relevant we might see a
new approach to upwarard scaling superconducting materials. More
broadly these ideas fall within the approach of materials by design,
where we are trying to design materials that target particular
function, in this case superconducting properties with highest
$T_c$.

We are grateful to I. Lukyanchuk and T. Wehling for useful discussions. This
work was supported by by US DoE BES and LDRD at Los Alamos.


\begin{thebibliography}{99}
\bibitem{comment2} One can deposit the Ca atoms on the single graphene
sheet. However in this case effect of top Ca atoms will be very
different and one would need a detail characterization of a
substrate.
\bibitem{Comment1} It might also be possible to cover bilayer graphene with
Ca, adding carriers and likely improving superconducting properties, but the
phonon properties of a Ca overlayer will be very different from that of of
intercalated Ca and are beyond the scope of this Letter.
\end{thebibliography}
\end{document}